\documentclass[prd,preprint,showpacs,superscriptaddress]{revtex4}
\usepackage{graphicx}

\begin{document}

\title{Production of Scalar Higgs Bosons Associated with $Z^0$ Boson\\
at the CERN LHC in the MSSM}

\author{Li Lin Yang}
\author{Chong Sheng Li}
\email{csli@pku.edu.cn}
\author{Jian Jun Liu}
\affiliation{Department of Physics, Peking University, Beijing
100871, China}
\author{Li Gang Jin}
\affiliation{Institute of Theoretical Physics, Academia Sinica,
  P. O. Box 2735, Beijing 100080, China}

\date{\today}

\begin{abstract}
We investigate the associated production of a scalar Higgs boson
($h^0$ or $H^0$) with $Z^0$ boson in the minimal supersymmetric
extension of the standard model (MSSM) at the CERN Large Hadron
Collider (LHC), including the contributions from $b\bar{b}$
annihilation at the tree level and gluon fusion via quark and
squark loops. We quantitatively analyze the total cross sections
in the mSUGRA scenario. For the production of $h^0$ associated
with $Z^0$, we find that in most of the parameter regions, the
contributions from initial $b\bar{b}$ and $gg$ are at a level of
one percent of the total cross section and therefore almost
insignificant. For the production of $H^0$ associated with $Z^0$,
the contributions from $b\bar{b}$ channel can be much larger than
those from light quark initial states. Especially for large
$\tan\beta$, the increment can reach about one order of magnitude.
Thus, when considering the associated production of $H^0$ and
$Z^0$ at the LHC, the contributions from $b\bar{b}$ annihilation
should be taken into account seriously.
\end{abstract}


\maketitle

\section{\label{sec:intro}Introduction}

The search for Higgs bosons is one of the main goals of the CERN
Large Hadron Collider (LHC), with $\sqrt{s}=14$TeV and a
luminosity of 100 fb$^{-1}$ per year \cite{lhc}. In the standard
model (SM), the Higgs boson mass is basically a free parameter
with an upper bound of $m_H\leq 600-800 $GeV \cite{massh}.
However, present data from precision measurements of electroweak
quantities indicate the existence of a light Higgs boson($m_H<204$
GeV at $95\%$ C.L.) and direct searches rule out the case
$m_H<114$GeV \cite{pdg}. In addition, in various extensions of the
SM, for example, in the two-Higgs-doublet models (THDM)
\cite{thdm}, particularly in the minimal supersymmetric standard
model (MSSM) \cite{mssm}, there are five physical Higgs particles:
two neutral CP-even bosons $h^0$ and $H^0$, one neutral CP-odd
boson $A^0$, and two charged bosons $H^\pm$. The Higgs boson $h^0$
should be lightest, with a mass $m_{h^0}\leq 135$GeV when
including the radiative corrections \cite{massh0}. It has been
shown \cite{detect} that the $h^0$ boson cannot escape detection
at the LHC and that in large areas of the parameter space, more
than one Higgs particle can be found.

At the LHC, the neutral Higgs bosons can be produced through
following mechanisms: gluon fusion $gg \rightarrow \phi$
\cite{gg2h,gg2hnlo,gg2hnnlo,gg2hresum}, weak boson fusion $qq
\rightarrow qqV^*V^* \rightarrow qqh^0/qqH^0$ \cite{vv2h},
associated production with weak bosons \cite{v2vh}, associated
production with a heavy quark-antiquark pair $gg,q\bar{q}
\rightarrow t\bar{t}\phi/b\bar{b}\phi$ \cite{tth} and pair
production \cite{hpair}. In this paper we focus our attentions on
the production of $h^0/H^0$ in association with $Z^0$ boson in the
MSSM. This is one of the main discovery channels of neutral Higgs
bosons at the Tevatron, with the Higgs decays to $b\bar{b}$ and
$Z^0$ decays leptonicly. The main backgrounds to this signal are
$Zb\bar{b}$ production, $ZZ$ production and top quark production
\cite{tevatron}. At the LHC, it can be used to detect the
``invisible'' Higgs which decays into the LSPs. The detailed
analysis of signals and backgrounds can be found in
Ref.~\cite{invisible}, where they found the ratio of signal to
background can as large as 12. In the SM, the Drell-Yan production
process (light $q\bar{q}$ annihilation) of Higgs associated with
$Z^0$ boson \cite{v2vh}, as well as the gluon fusion process
\cite{kniehl}, has been studied previously. In the MSSM, the
Drell-Yan contributions are related to the ones in the SM by an
overall coefficient, but the contributions of gluon fusion are
generally different in the two models due to the different
couplings and additional contributions from new channels.
Meanwhile, there are potentially important contributions to the
production of neutral Higgs bosons from $b\bar b$ annihilation at
the tree-level. In the MSSM, the Yukawa couplings $b$--$b$--$\phi$
can enhance Higgs boson production cross section via $b\bar b$
annihilation significantly for large $\tan\beta$. Therefore,
besides the production channel via Drell-Yan process, in order to
obtain the complete leading order (LO) total cross sections for
the associated production of scalar Higgs bosons with $Z^0$ boson,
the partonic subprocesses $b\bar{b}\rightarrow Zh^0/ZH^0$ should
also be taken into account. Moreover, the loop induced subprocess
$gg \rightarrow Zh^0/ZH^0$ become important due to the potential
enhancement, and should be considered, too. We will investigate
all these contributions in our work.

Our paper is organized as follows. In section \ref{sec:cal}, we
list analytical results for the tree level cross sections of $pp
\rightarrow b\bar b \rightarrow Zh^0/ZH^0$ and loop induced cross
sections of $pp \rightarrow gg \rightarrow Zh^0/ZH^0$ in the MSSM.
In section \ref{sec:num}, we present quantitative predictions for
the inclusive cross section of $pp \rightarrow Zh^0/ZH^0+X$ at the
LHC adopting the MSSM parameters constrained within the minimal
supergravity (mSUGRA) scenario and discuss the implications of our
results. The relevant MSSM couplings and form factors are given in
the Appendix \ref{sec:cp} and \ref{sec:ff}, respectively.

\section{\label{sec:cal}Calculation}

\begin{figure}
\begin{center}
\includegraphics{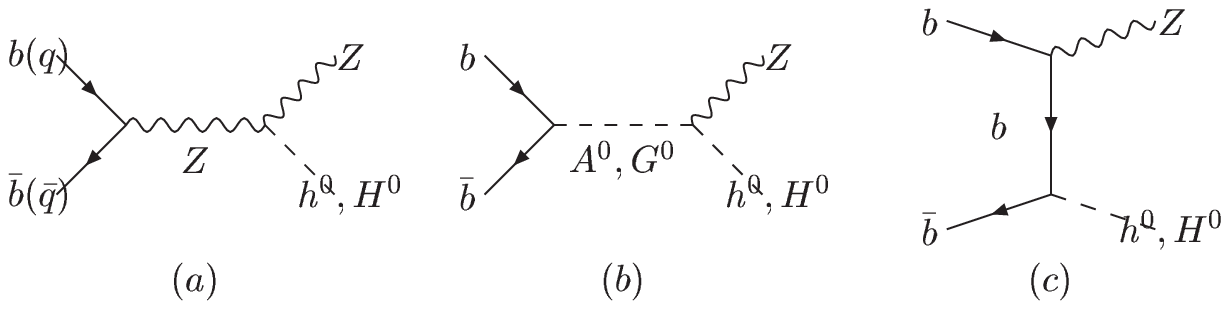}
\end{center}
\caption{\label{fig:feyn1}
Feynman diagrams for the subprocess $b\bar{b} \rightarrow ZH$}
\end{figure}

\begin{figure}
\begin{center}
\includegraphics{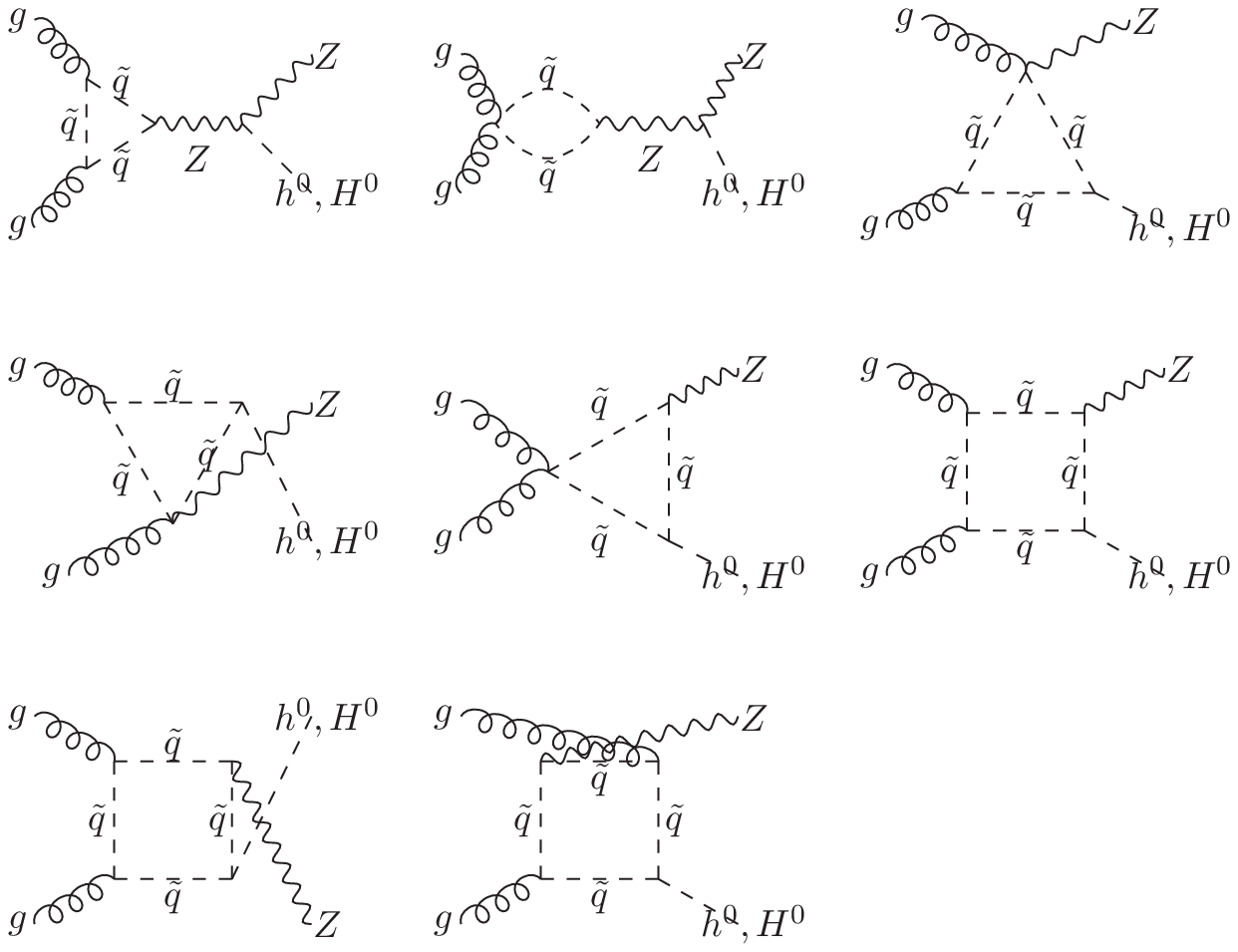}
\end{center}
\caption{\label{fig:feyn3} Feynman diagrams for the subprocess $gg
\rightarrow ZH$ (including only squark loop)}
\end{figure}

The relevant Feynman diagrams are created by FeynArts
\cite{feynarts} version 3.2 automatically and are shown in
Fig.~\ref{fig:feyn1}--\ref{fig:feyn3}. We carry out the
calculation in the 't~Hooft--Feynman gauge and use dimensional
reduction for regularization of the ultraviolet divergences in the
loop diagrams. In the following expressions, $G_{L,R}^{\bar{f}fZ}$
and $G^{ijk}$ are the couplings, which are given explicitly in
Appendix \ref{sec:cp}; $H$ stands for the scalar Higgs bosons,
$h^0$ or $H^0$; $S$ and $f$ are propagating scalar and fermion
particles, respectively.

For the partonic subprocesses
\begin{eqnarray*}
b(k_1) + \bar{b}(k_2) &\rightarrow& Z(k_3, \varepsilon_3) + H(k_4), \\
g(k_1, \varepsilon_1) + g(k_2, \varepsilon_2) &\rightarrow&
 Z(k_3, \varepsilon_3) + H(k_4),
\end{eqnarray*}
we define the Mandelstam variables as
\begin{equation}
\hat{s}=(k_1+k_2)^2, \quad \hat{t}=(k_1-k_3)^2, \quad
\hat{u}=(k_1-k_4)^2.
\end{equation}

The tree-level amplitude $\mathcal{M}^{b\bar{b}}$ for the
subprocess $b\bar{b} \rightarrow ZH$ consists of the three
diagrams $(a)-(c)$ in Fig.~\ref{fig:feyn1}. The previous results
\cite{v2vh} for vector boson bremsstrahlung include only the first
diagram, where the $b\bar{b}$ pair replaced by light
quark-antiquark pairs. We also recompute the contributions of
light quarks to the tree-level cross sections and compare them
with ones of $b\bar{b} \rightarrow ZH$.

Using the notations defined above, the amplitude
$\mathcal{M}^{b\bar{b}}$ can be expressed as
\begin{equation}
\mathcal{M}^{b\bar{b}}= \mathcal{M}^{b\bar{b}}_a +
\mathcal{M}^{b\bar{b}}_b + \mathcal{M}^{b\bar{b}}_c
\end{equation}
with
\begin{eqnarray}
\mathcal{M}^{b\bar{b}}_a &=& \bar{v}(k_2)\not{\varepsilon_3}
(G_L^{\bar{b}bZ}P_L + G_R^{\bar{b}bZ}P_R)u(k_1)
\frac{-iG^{HZZ}}{\hat{s}-m_Z^2}, \\
\mathcal{M}^{b\bar{b}}_b &=& \sum_{S=A^0,G^0}
\bar{v}(k_2)G^{b\bar{b}S}\gamma^5u(k_1)
\frac{i G^{HSZ}}{\hat{s}-m_S^2+im_S\Gamma_S}(k_3+2k_4)\cdot \varepsilon_3,\\
\mathcal{M}^{b\bar{b}}_c &=& \bar{v}(k_2)G^{b\bar{b}H}
\frac{i(\not{k_1}-\not{k_3}+m_b)}{\hat{t}-m_b^2}\not{\varepsilon_3}
(G_L^{\bar{b}bZ}P_L + G_R^{\bar{b}bZ}P_R)u(k_1) \nonumber \\
&+& \bar{v}(k_2)\not{\varepsilon_3}(G_L^{\bar{b}bZ}P_L + G_R^{\bar{b}bZ}P_R)
\frac{i(\not{k_1}-\not{k_4}+m_b)}{\hat{u}-m_b^2}
G^{b\bar{b}H}u(k_1).
\end{eqnarray}
Here $P_{L,R}\equiv(1\mp\gamma^5)/2$, $m_{G^0}=m_Z$, $\Gamma_{G^0}=0$,
and $\Gamma_{A^0}$ is the decay width of $A^0$.

The gluon fusion subprocess is forbidden at tree-level. At
one-loop level, in general, the cross section will receive
contributions from both quark loops and squark loops,
as shown in Fig.~\ref{fig:feyn2}, \ref{fig:feyn3}. Note that each
diagram actually represents a couple of diagrams with opposite
directions of charge flow. However, we find that the contributions
from each pair of squark loop diagrams to this process cancel each
other due to the opposite signs of momenta. So the gluon fusion
cross section arises only from the quark loop diagrams, i.e.,
\begin{equation}
\mathcal{M}^{gg}=
\mathcal{M}^{gg}_a+\mathcal{M}^{gg}_b+\cdots+\mathcal{M}^{gg}_e,
\end{equation}
where the subscripts $a-e$ refer to the corresponding diagrams in
Fig.~\ref{fig:feyn2}.

In general, the amplitudes $\mathcal{M}^{gg}$ can be written as a linear
combination of the invariants formed by three independent external momenta
$k_1, k_2, k_3$, polarization vectors of three external gauge bosons
$\varepsilon_1, \varepsilon_2, \varepsilon_3$,
metric tensor $g^{\mu\nu}$ and Levi-Civita tensor
$\epsilon^{\mu\nu\rho\sigma}$,
in which the terms without Levi-Civita tensor vanish.
Thus, with taking into account the relations
$\varepsilon_1 \cdot k_1 = \varepsilon_2 \cdot k_2 =
\varepsilon_3 \cdot k_3 = 0$, 24 terms remain in the amplitudes.
It is easy to prove the following identity
\begin{eqnarray}
g^{\mu\nu}\epsilon^{\rho\sigma\alpha\beta}
&-& g^{\mu\rho}\epsilon^{\nu\sigma\alpha\beta}
 - g^{\mu\alpha}\epsilon^{\nu\rho\sigma\beta}
 + g^{\mu\beta}\epsilon^{\nu\rho\sigma\alpha} \\ \nonumber
&+& g^{\nu\sigma}\epsilon^{\mu\rho\alpha\beta}
 - g^{\rho\sigma}\epsilon^{\mu\nu\alpha\beta}
 + g^{\sigma\alpha}\epsilon^{\mu\nu\rho\beta}
 - g^{\sigma\beta}\epsilon^{\mu\nu\rho\alpha} = 0.
\end{eqnarray}
Using it, one can immediately see that not all of the 24 terms are
linear independent.
Actually, there are only 14 independent ones,
and finally the amplitudes can be written as
\begin{eqnarray}
\nonumber
\mathcal{M}^{gg}&=&
A_1 \epsilon^{\varepsilon_1\varepsilon_2\varepsilon_3k_1}
+A_2 \epsilon^{\varepsilon_1\varepsilon_2\varepsilon_3k_2}
+A_3 \epsilon^{\varepsilon_1\varepsilon_2\varepsilon_3k_3}
+A_4 \epsilon^{\varepsilon_1\varepsilon_2k_1k_2} \varepsilon_3 \cdot k_1
+A_5 \epsilon^{\varepsilon_1\varepsilon_2k_1k_2} \varepsilon_3 \cdot k_2 \\
\nonumber
&& +A_6 \epsilon^{\varepsilon_1\varepsilon_3k_1k_3} \varepsilon_2 \cdot k_1
+A_7 \epsilon^{\varepsilon_1\varepsilon_3k_1k_3} \varepsilon_2 \cdot k_3
+A_8 \epsilon^{\varepsilon_2\varepsilon_3k_2k_3} \varepsilon_1 \cdot k_2
+A_9 \epsilon^{\varepsilon_2\varepsilon_3k_2k_3} \varepsilon_1 \cdot k_3 \\
&& +A_{10} \epsilon^{\varepsilon_1k_1k_2k_3} \varepsilon_2 \cdot \varepsilon_3
+A_{11} \epsilon^{\varepsilon_2k_1k_2k_3} \varepsilon_1 \cdot \varepsilon_3
+A_{12} \epsilon^{\varepsilon_3k_1k_2k_3} \varepsilon_1 \cdot \varepsilon_2 \\
\nonumber
&& +A_{13} \epsilon^{\varepsilon_1\varepsilon_2k_2k_3} \varepsilon_3 \cdot k_1
+A_{14} \epsilon^{\varepsilon_1\varepsilon_2k_1k_3} \varepsilon_3 \cdot k_2.
\end{eqnarray}
The explicit expressions of the form factors $A_i(i=1,...,14)$
are shown in Appendix \ref{sec:ff}.

The differential cross sections of the subprocesses are given by
\begin{equation}
\frac{d\hat{\sigma}}{d\hat{t}}=
\frac{1}{16\pi\hat{s}(\hat{s}-4m^2)}
\overline{\sum}|\mathcal{M}|^2,
\end{equation}
where $\mathcal{M}=\mathcal{M}^{b\bar{b}}, m=m_b$ for $b\bar{b}$ channel
and $\mathcal{M}=\mathcal{M}^{gg}, m=0$ for gluon fusion.

Using the standard factorization procedure, the total cross section can be
obtained as a convolution of partonic cross sections with corresponding parton
distribution functions,
\begin{equation}
\sigma(pp \rightarrow ZH)=\sum_{\alpha,\beta} \frac{1}{1+\delta_{\alpha\beta}}
\int_{\tau_0}^1 d x_1 \int_{\tau_0/x_1}^1 d x_2
\left[f_{\alpha/p}(x_1) f_{\beta/p}(x_2)+(\alpha \leftrightarrow \beta)\right]
\hat{\sigma}(\alpha\beta \rightarrow ZH),
\end{equation}
where $\tau_0=(m_Z+m_H)^2/s$, $\sqrt{s}$ is the center-of-mass energy of
the LHC,
$\alpha$ and $\beta$ denote the initial partons.

\section{\label{sec:num}Numerical Results and Conclusions}

In the following we present some numerical results. In our
numerical calculations, the SM input parameters were taken to be
$\alpha(m_Z) = 1/128.8$, $m_W=80.423$GeV, $m_Z=91.188$GeV and
$m_t=174.3$GeV \cite{pdg}. We use the one-loop evolution of the
strong coupling constant $\alpha_s(Q)$ \cite{alphas,spira} with
$\alpha_s(m_Z) = 0.1172$. The leading order CTEQ6 parton
distribution functions \cite{cteq} are used here and the
factorization scale is taken to be the invariant mass of the two
final particles $\mu=m_{ZH}=\sqrt{(p_Z+p_H)^2}$. Moreover, in
order to improve the perturbative calculations, we take the
running mass of bottom quark $m_b(Q)$ evaluated by the one-loop
formula
\begin{equation}
m_b(Q) = U_6(Q,m_t)U_5(m_t,m_b)m_b(m_b),
\end{equation}
with $m_b(m_b)=4.25$GeV.
The evolution factor $U_f$ is
\begin{equation}
U_f(Q_2,Q_1) = \left( \frac{\alpha_s(Q_2)}{\alpha_s(Q_1)} \right)^{d_f}
\nonumber
\end{equation}
with
\begin{equation}
d_f = \frac{12}{33-2f}.
\nonumber
\end{equation}
The relevant MSSM parameters, the Higgs masses and mixing angles
$\alpha$, are determined in the mSUGRA scenario as implemented in
program package ISAJET 7.69 \cite{isajet}. The GUT parameters
$m_0$, $A_0$ and $\mathrm{sgn}(\mu)$ were taken to be
$m_0=200$~GeV, $A_0=-100$~GeV, $\mu>0$. $m_{1/2}$, $\tan\beta$
were varied to obtain various Higgs masses. Actually, in mSUGRA
scenario, the mass of the light scalar Higgs boson $h^0$ can only
reach about 125 GeV, which is independent of the choice of GUT
parameters. It should be noted that in some parameter region,
$m_{A^0}
> m_Z + m_{h^0}$, and the momentum of $A^0$ can approach its mass shell,
which will lead to a singularity arising from the $A^0$
propagator. This can be avoided by introduing the non-zero decay
width $\Gamma_{A^0}$, which was also calculated by ISAJET.

\begin{figure}[ht]
\begin{center}
\begin{tabular}{c}
\includegraphics[width=0.7\textwidth]{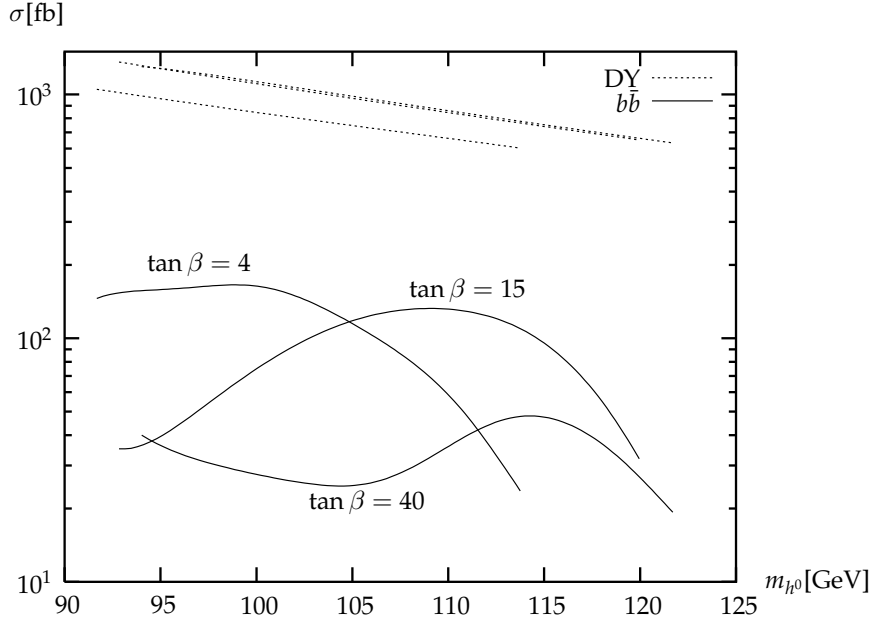} \\
(a) \\
\includegraphics[width=0.7\textwidth]{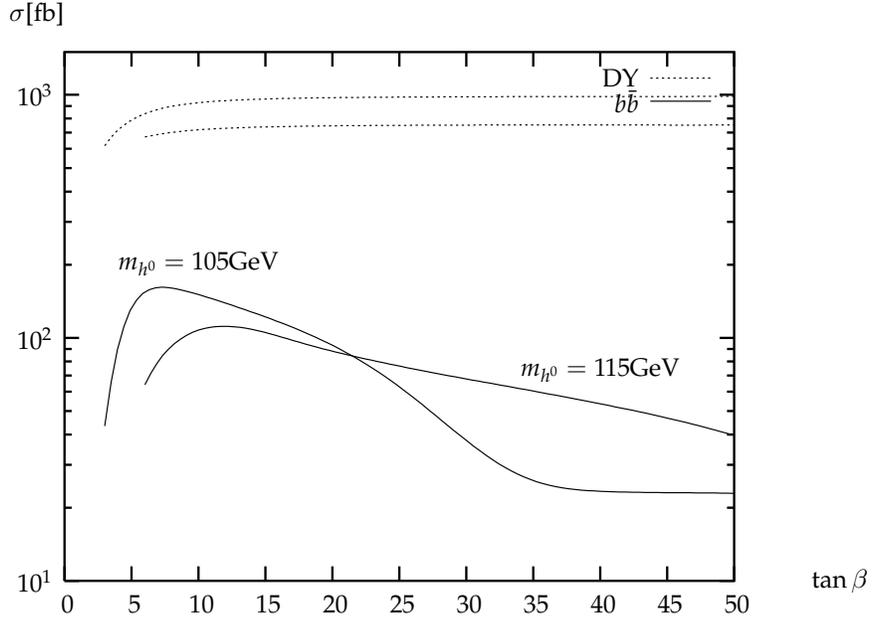} \\
(b)
\end{tabular}
\end{center}
\caption{\label{fig:num1}
Total cross sections $\sigma$ (in fb) of $pp \rightarrow Zh^0$
via $b\bar{b}$ annihilation (solid lines) compared with Drell-Yan
ones (dotted lines) at the LHC
(a) as functions of $m_{h^0}$ for $\tan\beta=4$, 15 and 40;
and (b) as functions of $\tan\beta$ for $m_{h^0}=105$ and 115 GeV.}
\end{figure}

\begin{figure}[ht]
\begin{center}
\begin{tabular}{c}
\includegraphics[width=0.7\textwidth]{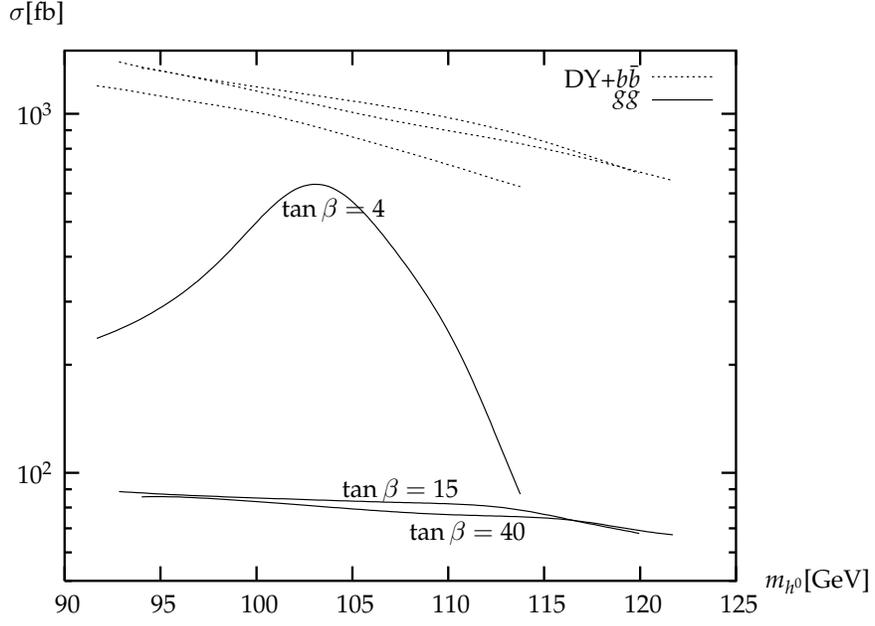} \\
(a) \\
\includegraphics[width=0.7\textwidth]{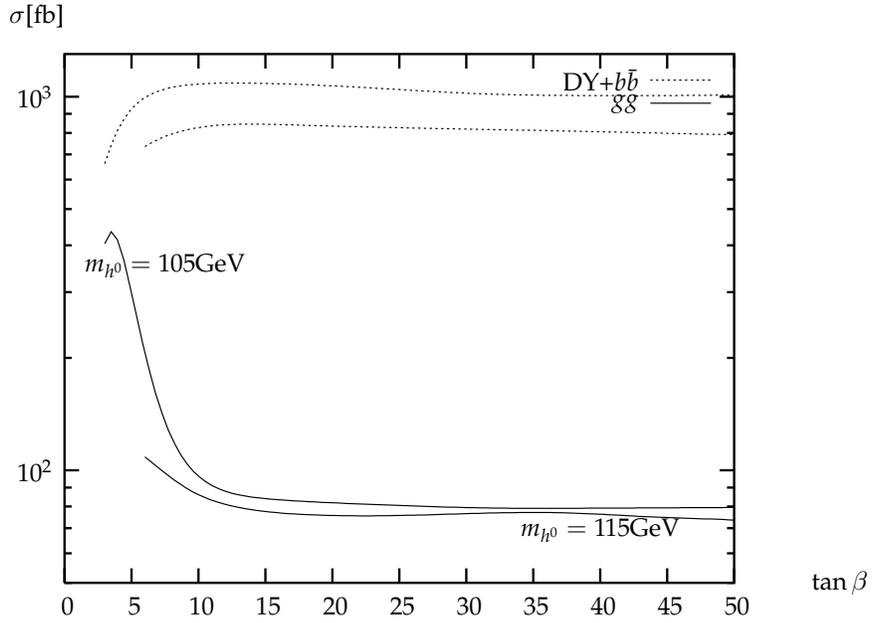} \\
(b)
\end{tabular}
\end{center}
\caption{\label{fig:num2}
Total cross section $\sigma$ (in fb) of $pp \rightarrow Zh^0$
via gluon fusion (solid lines) compared with complete $q\bar{q}$
annihilation ones (dotted lines) at the LHC
(a) as functions of $m_{h^0}$ for $\tan\beta=4$, 15 and 40;
and (b) as functions of $\tan\beta$ for $m_{h^0}=105$ and 115 GeV.}
\end{figure}

Fig.~\ref{fig:num1}$(a)$ and \ref{fig:num2}$(a)$ show the total
cross sections for the process $pp \rightarrow Zh^0 + X$ versus
the light scalar Higgs boson mass $m_{h^0}$ for $\tan\beta=4,15$
and $40$. Fig.~\ref{fig:num1}$(b)$ and \ref{fig:num2}$(b)$ show
the dependence of the cross sections on $\tan\beta$ for
$m_{h^0}=105$ and $115$~GeV. From Fig.~\ref{fig:num1} we can see
that the $b\bar{b}$ contributions are approximately one order of
magnitude smaller than the ones of Drell-Yan process \cite{v2vh},
and only increase the total cross sections by about several
percents, which are smaller than the QCD corrections to the
Drell-Yan process. Fig.~\ref{fig:num2} shows the contributions of
gluon fusion to the cross sections. Comparing with ones of the
complete $q\bar{q}$ annihilation (i.e. Drell-Yan + $b\bar{b}$),
one can see that for a light $h^0$ (for example,
$m_{h^0}=105$~GeV), the contributions of gluon fusion are about
half of the ones of the $q\bar{q}$ annihilation for low
$\tan\beta(\le 5)$, but with the increasing of $\tan\beta$ from 4
to 10, the former decrease significantly, and become one order of
magnitude smaller than the latter in general. Therefore, in most of
the parameter regions, the contributions of both $b\bar{b}$ and $gg$
initial states are almost insignificant and can be neglected.

\begin{figure}[ht]
\begin{center}
\begin{tabular}{c}
\includegraphics[width=0.7\textwidth]{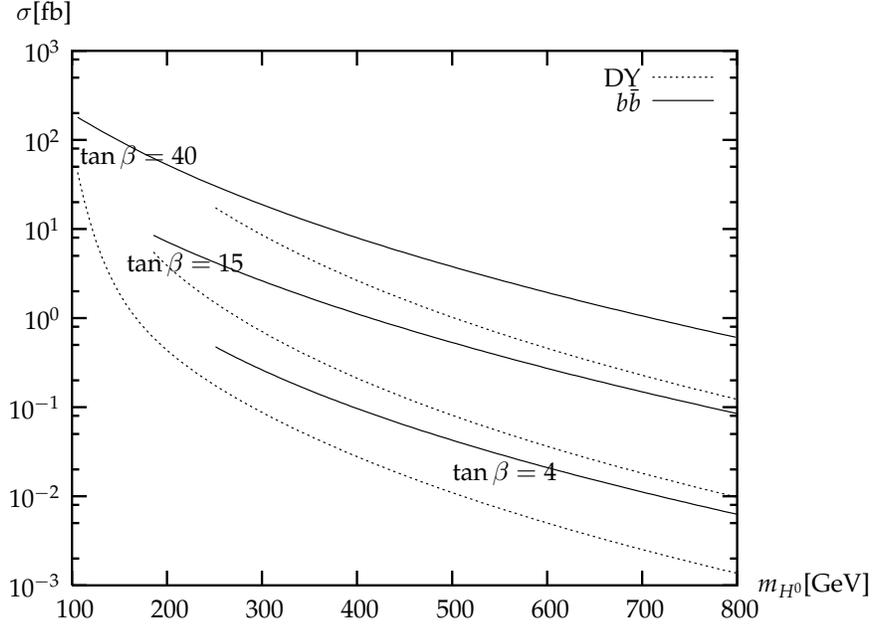} \\
(a) \\
\includegraphics[width=0.7\textwidth]{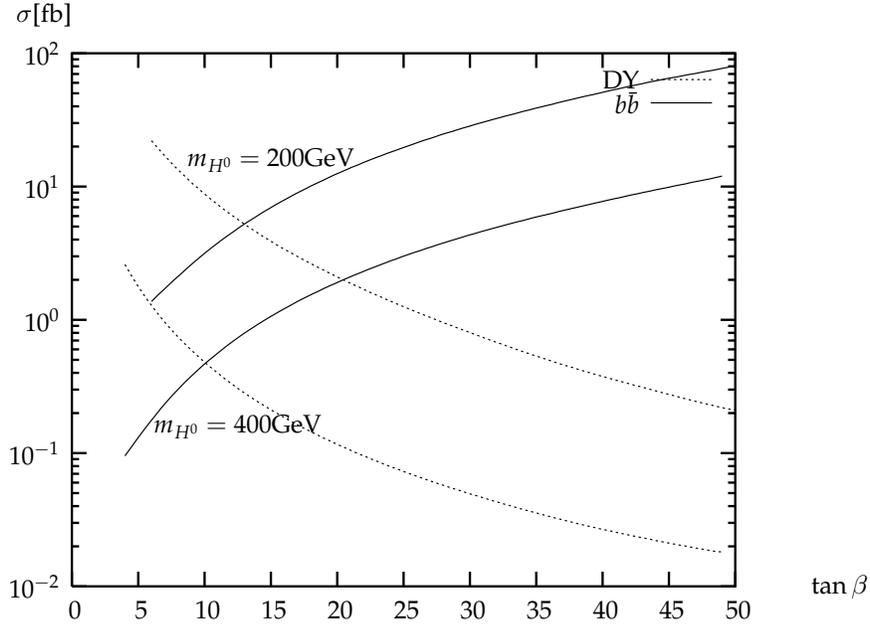} \\
(b)
\end{tabular}
\end{center}
\caption{\label{fig:num3}
Total cross sections $\sigma$ (in fb) of $pp \rightarrow ZH^0$
via $b\bar{b}$ annihilation (solid lines) compared with Drell-Yan
ones (dotted lines) at the LHC
(a) as functions of $m_{H^0}$ for $\tan\beta=$ 4 (curves starting from
$m_{H^0}$=250GeV, 15 (curves starting from $m_{H^0}$=200GeV and
40 (curves starting from $m_{H^0}$=100GeV);
and (b) as functions of $\tan\beta$ for $m_{H^0}=200$ and 400 GeV.}
\end{figure}

\begin{figure}[ht]
\begin{center}
\begin{tabular}{c}
\includegraphics[width=0.7\textwidth]{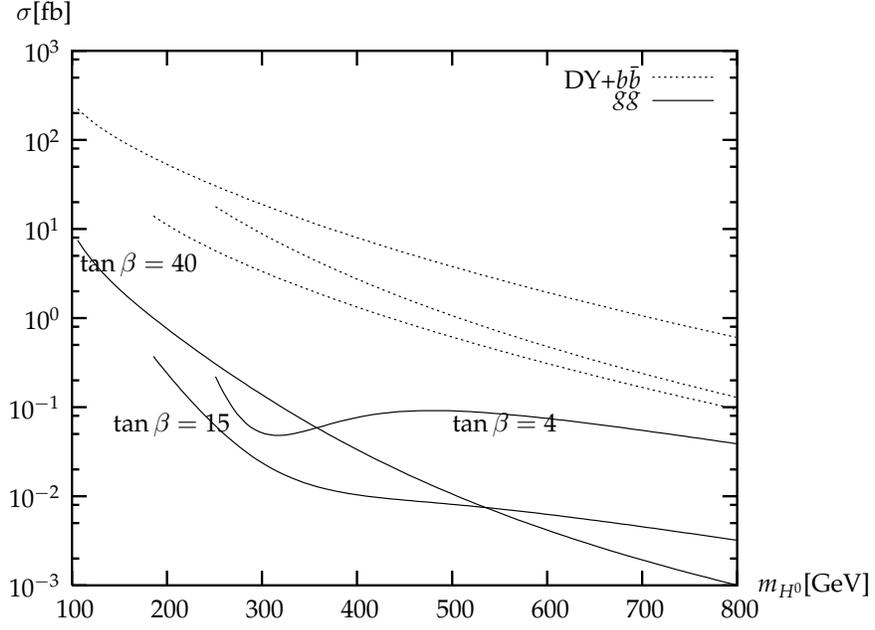} \\
(a) \\
\includegraphics[width=0.7\textwidth]{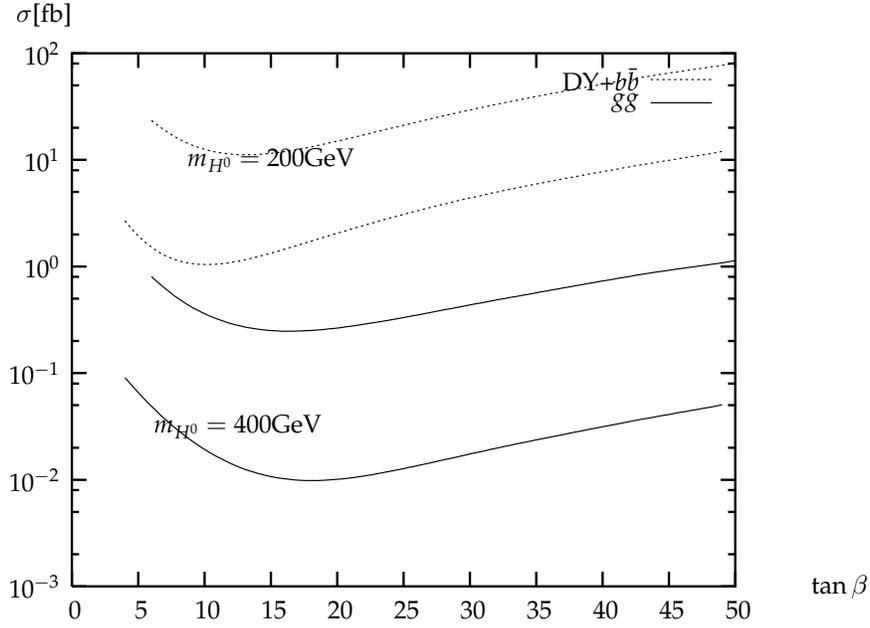} \\
(b)
\end{tabular}
\end{center}
\caption{\label{fig:num4}
Total cross section $\sigma$ (in fb) of $pp \rightarrow ZH^0$
via gluon fusion (solid lines) compared with complete $q\bar{q}$
annihilation ones (dotted lines) at the LHC
(a) as functions of $m_{H^0}$ for $\tan\beta=4$, 15 and 40(the mass ranges
are the same as Fig.~\ref{fig:num3});
and (b) as functions of $\tan\beta$ for $m_{H^0}=200$ and 400 GeV.}
\end{figure}

Fig.~\ref{fig:num3}$(a)$ and \ref{fig:num4}$(a)$ show the total
cross sections for the process $pp \rightarrow ZH^0 + X$ as a
function of the heavy scalar Higgs boson mass $m_{H^0}$ for the
three representative values of $\tan\beta$.
Fig.~\ref{fig:num3}$(b)$ and \ref{fig:num4}$(b)$ show the
dependence of the cross sections on $\tan\beta$ for $m_{H^0}=200$
and $400$~GeV. We found that the $b\bar{b}$ contributions to the
total cross sections for the production of heavier neutral Higgs
boson can be much larger than the contributions of Drell-Yan
process for large $\tan\beta$, as shown in Fig.~\ref{fig:num3}.
For example, when $\tan\beta=40$ and $m_{H^0}=200$~GeV, the
$b\bar{b}$ cross section is about $50$~fb, while one of Drell-Yan
process is less than $0.5$~fb. Therefore, the $b\bar{b}$ channel
provides a dramatic enhancement to the total cross sections, and
should definitely be considered. On the other hand, the
contributions of gluon fusion are very small ($<1$~fb) and can be
neglected in most cases, as shown in Fig.~\ref{fig:num4}.

\begin{figure}[ht]
\begin{center}
\includegraphics[width=0.7\textwidth]{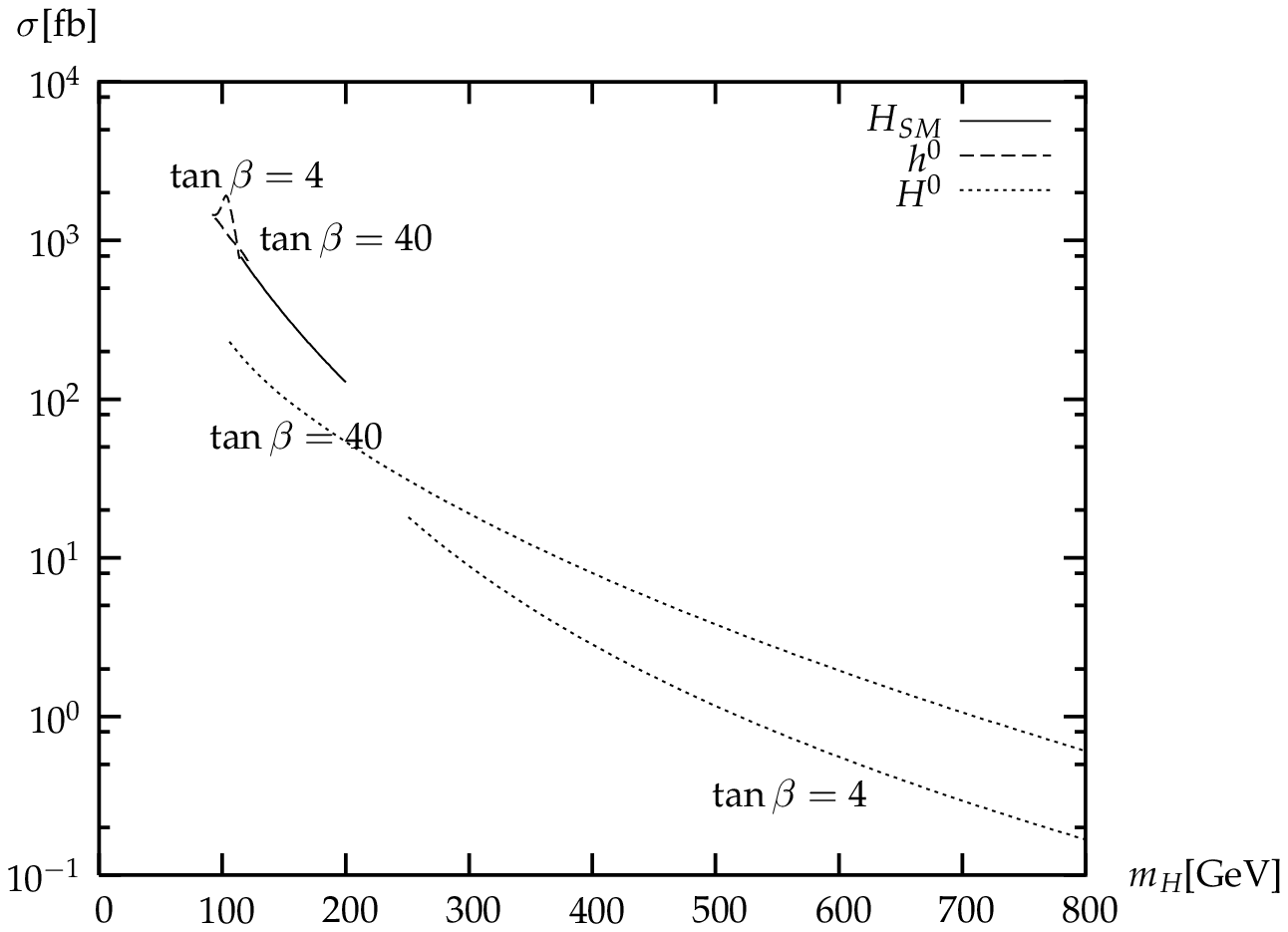}
\end{center}
\caption{\label{fig:num6}A comparison of the production cross sections of
the three Higgs bosons as functions of their masses.}
\end{figure}

For the comparison of the production rates of $H_{SM}$, $h^0$ and
$H^0$, they are displayed as the functions of their masses in
Fig.~\ref{fig:num6}, where all the contributions (i.e. Drell-Yan +
$b\bar{b}$ + gluon fusion) have been included for $\tan\beta=4$
and 40, respectively. For the SM Higgs, as shown in
Ref.~\cite{kniehl}, the gluon fusion contributions are negligible
for Higgs masses allowed by experiments, so we didn't include them
here. From Fig.~\ref{fig:num6}, one can find that no matter what
the value of $\tan\beta$ is, the production rates of $H^0$ are the
smallest, the ones of $h^0$ are the largest, and the ones of
$H_{SM}$ are medium. This feature indicates that the predictions
for the associated production of the Higgs bosons and $Z^0$ boson
in the SM and the MSSM are different quantitatively and
distinguishable, which is in agreement with the results shown in
Ref.~\cite{spira}, where only Drell-Yan contributions are
included.

In conclusion, we have calculated the scalar Higgs bosons
($h^0,H^0$) production in association with a $Z^0$ boson through
both $b\bar{b}$ channel and gluon fusion in the mSUGRA model at
the LHC. Our results show that in most of the parameter regions,
the contributions to the total cross section for associated
production of the light scalar Higgs boson $h^0$ and $Z^0$ boson
mainly come from light quark annihilation and the contributions
from initial $b\bar{b}$ and $gg$ are small and negligible. And the
total cross section for associated production of heavy scalar
Higgs boson $H^0$ and $Z^0$ boson, with including the
contributions of $b\bar{b}$ channel, can be increased greatly.
Especially for large $\tan\beta$, such increment can reach about
one order of magnitude. Thus, the contributions of $b\bar{b}$
channel should definitely be taken into account, and the ratio of
signal to background obtained in Ref.~\cite{invisible}, which is
based on the previous calculations, can be enhanced at large
$\tan\beta$. The contributions from gluon fusion are still small
and unimportant.

\begin{acknowledgments}
This work was supported in part by the National Natural Science
Foundation of China and Specialized Research Fund for the Doctoral
Program of Higher Education .
\end{acknowledgments}

\appendix

\section{\label{sec:cp}Couplings}

Here we list the relevant couplings in the
amplitudes. $i, j$ stand for generation indices and
$r, s$ stand for color indices.
\begin{displaymath}
\begin{array}{ll}
\displaystyle{G_L^{\bar{u}^i u^j Z} = (-3+4s_W^2)\frac{ie\delta^{ij}}{6c_Ws_W}},
&
\displaystyle{G_R^{\bar{u}^i u^j Z} = \frac{2ies_W\delta^{ij}}{3c_W}},
\\
\displaystyle{G_L^{\bar{d}^i d^j Z} = -(-3+2s_W^2)\frac{ie\delta^{ij}}{6c_Ws_W}},
&
\displaystyle{G_R^{\bar{d}^i d^j Z} = -\frac{ies_W\delta^{ij}}{3c_W}},
\\
\displaystyle{G^{h^0ZZ} = \frac{ieM_Ws_{\beta-\alpha}}{c_W^2s_W}},
&
\displaystyle{G^{H^0ZZ} = \frac{ieM_Wc_{\beta-\alpha}}{c_W^2s_W}},
\\
\displaystyle{G^{u^i \bar{u}^j h^0} = -\frac{i c_\alpha e \delta^{ij} m_{u^i}}
{2M_W s_\beta s_W}},
&
\displaystyle{G^{d^i \bar{d}^j h^0} = \frac{i e s_\alpha \delta^{ij} m_{d^i}}
{2c_\beta M_W s_W}},
\\
\displaystyle{G^{u^i \bar{u}^j H^0} = -\frac{i e s_\alpha \delta^{ij} m_{u^i}}
{2M_W s_\beta s_W}},
&
\displaystyle{G^{d^i \bar{d}^j H^0} = -\frac{i c_\alpha e \delta^{ij} m_{d^i}}
{2c_\beta M_W s_W}},
\\
\displaystyle{G^{u^i \bar{u}^j A^0} = -\frac{e\delta^{ij}m_{u^i}}{2M_Ws_Wt_\beta}},
&
\displaystyle{G^{d^i \bar{d}^j A^0} = -\frac{et_\beta\delta^{ij}m_{d^i}}{2M_Ws_W}},
\\
\displaystyle{G^{u^i \bar{u}^j G^0} = -\frac{e\delta^{ij}m_{u^i}}{2M_Ws_W}},
&
\displaystyle{G^{d^i \bar{d}^j G^0} = \frac{e\delta^{ij}m_{d^i}}{2M_Ws_W}},
\\
\displaystyle{G^{h^0A^0Z} = \frac{c_{\beta-\alpha}e}{2c_Ws_W}},
&
\displaystyle{G^{h^0G^0Z} = \frac{es_{\beta-\alpha}}{2c_Ws_W}},
\\
\displaystyle{G^{H^0A^0Z} = -\frac{es_{\beta-\alpha}}{2c_Ws_W}},
&
\displaystyle{G^{H^0G^0Z} = \frac{c_{\beta-\alpha}e}{2c_Ws_W}},
\\
\displaystyle{G^{\bar{u}^i_ru^j_sg^a} = -ig_s\delta^{ij}T^a_{rs}},
&
\displaystyle{G^{\bar{d}^i_rd^j_sg^a} = -ig_s\delta^{ij}T^a_{rs}}.
\end{array}
\end{displaymath}
with
\begin{eqnarray*}
s_W = \sin\theta_W,\, c_W = \cos\theta_W,\, s_\alpha=\sin\alpha,\,
c_\alpha = \cos\alpha,\, s_\beta = \sin\beta, \\
c_\beta = \cos\beta,\, t_\beta = \tan\beta,\, s_{\beta-\alpha} =
\sin(\beta-\alpha),\, c_{\beta-\alpha} = \cos(\beta-\alpha).
\end{eqnarray*}

\section{\label{sec:ff}Form Factors}

\begin{figure}
\begin{center}
\includegraphics{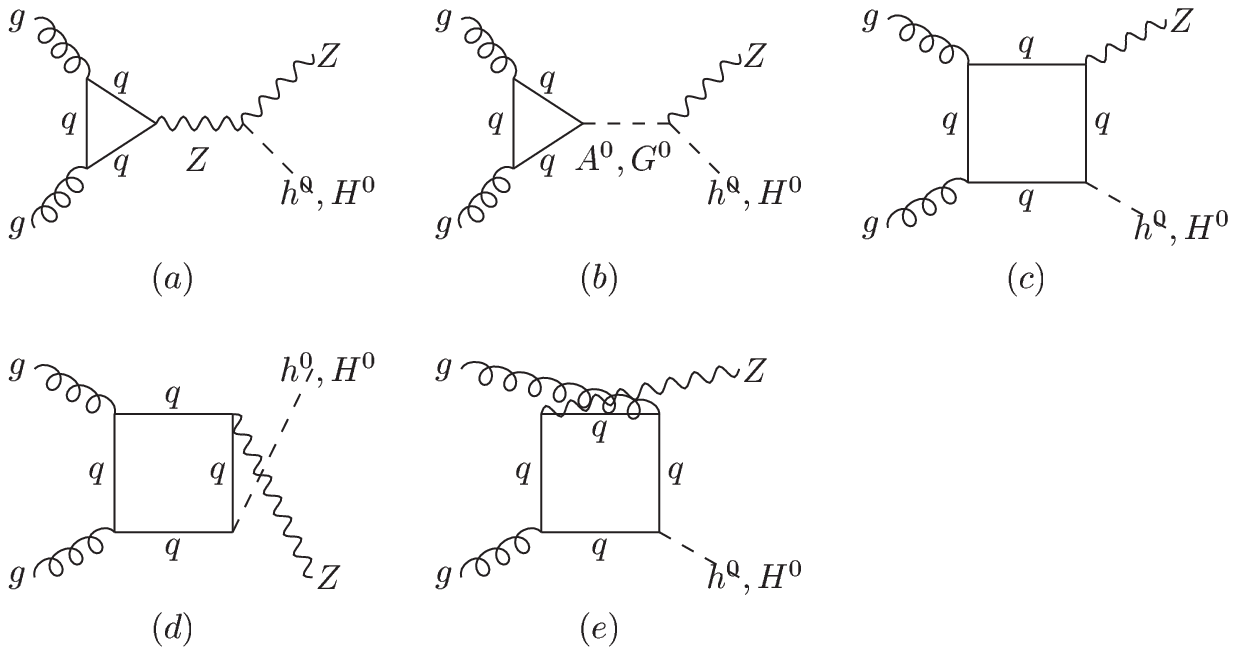}
\end{center}
\caption{\label{fig:feyn2}
Feynman diagrams for the subprocess $gg \rightarrow ZH$
(including only quark loop)}
\end{figure}

This appendix lists all the coefficient $A$s in the amplitude of
subprocess $gg \rightarrow ZH$, in terms of 3- and 4-points one-loop
integrals\cite{loop}.
The diagrams $(a)-(e)$ refer to those in Fig. \ref{fig:feyn2}.
For convenience, we define abbreviations of one-loop integrals
for each diagram as following,
\begin{eqnarray*}
C^{(a)} &=& C^{(b)} = C(0, \hat{s}, 0, m_f^2, m_f^2, m_f^2) \\
C^{(c)} &=& C(m_H^2, m_Z^2, \hat{s}, m_f^2, m_f^2, m_f^2) \\
D^{(c)} &=& D(0, m_H^2, m_Z^2, 0, \hat{t}, \hat{s},
m_f^2, m_f^2, m_f^2, m_f^2) \\
C^{(d)} &=& C(m_Z^2, m_H^2, \hat{s}, m_f^2, m_f^2, m_f^2) \\
D^{(d)} &=& D(0, m_Z^2, m_H^2, 0, \hat{u}, \hat{s},
m_f^2, m_f^2, m_f^2, m_f^2) \\
C^{(e)} &=& C(0, m_H^2, \hat{t}, m_f^2, m_f^2, m_f^2) \\
D^{(e)} &=& D(m_Z^2, 0, m_H^2, 0, \hat{u}, \hat{t},
 m_f^2, m_f^2, m_f^2, m_f^2)
\end{eqnarray*}
Note that in the following expressions there is an implicit sum over $f$
for $f=t,b$.

For diagram $(a)$ and $(b)$, there are only two coefficients
which are not zero, respectively.
\begin{eqnarray*}
A_4^{(a)}&=&A_5^{(a)}=-\frac{1}{2\pi^2}G^{HZZ}(G_L^{\bar{f}fZ}-G_R^{\bar{f}fZ})
G^{\bar{f}fg_1}G^{\bar{f}fg_2}\frac{1}{\hat{s}-m_Z^2}C_{12}^{(a)} \\
A_4^{(b)}&=&A_5^{(b)}=\frac{1}{2\pi^2}G^{HSZ}(G_L^{f\bar{f}S}-G_R^{f\bar{f}S})
G^{\bar{f}fg_1}G^{\bar{f}fg_2}\frac{m_f}{\hat{s}-m_S^2+im_S\Gamma_S}C_0^{(b)}
\end{eqnarray*}

For diagram $(c)$, the coefficients are the following expressions
time an overall factor
$\frac{1}{8\pi^2}G^{f\bar{f}H}(G_L^{\bar{f}fZ} - G_R^{\bar{f}fZ})
G^{\bar{f}fg_1}G^{\bar{f}fg_2}m_f$
\begin{eqnarray*}
A_{1} &=& 4 \left[ C_0^{(c)} + C_1^{(c)} + C_2^{(c)} - 2 D_{00}^{(c)}
 + \hat{s}(D_{12}^{(c)} + D_{13}^{(c)}) \right]\\
 && \hspace{3cm} + (\hat{u} - m_Z^2) \left[ D_0^{(c)} - 2D_2^{(c)}
 + 4(D_{22}^{(c)} + D_{23}^{(c)}) \right] \\
A_{ 2} &=& 2C_0^{(c)} + 4(C_1^{(c)} + C_2^{(c)}) + (\hat{t} - m_Z^2)D_0^{(c)}
 - 2\hat{s} D_1^{(c)} \\
 && \hspace{3cm} + 8 D_{00}^{(c)}
 + 4(m_Z^2 + \hat{s} - \hat{t})D_{12}^{(c)} + 4\hat{s} D_{13}^{(c)} \\
A_3 &=& -2C_0^{(c)} - 4C_1^{(c)} + \hat{s} \left[ D_0^{(c)} - 2D_2^{(c)}
 - 4(D_{12}^{(c)} + D_{22}^{(c)} + D_{23}^{(c)}) \right] \\
A_4 &=& -4 \left[ D_0^{(c)} + D_2^{(c)} + D_3^{(c)} - 2(D_{12}^{(c)}
 + D_{13}^{(c)}) \right] \\
A_5 &=& 4 \left[ D_1^{(c)} + 2(D_{12}^{(c)} + D_{13}^{(c)}) \right] \\
A_6 &=& 4 \left[ D_2^{(c)} + D_3^{(c)} + 2(D_{22}^{(c)}
 + D_{23}^{(c)}) \right] \\
A_7 &=& A_9 = -4(D_2^{(c)} + 2D_{22}^{(c)}) \\
A_8 &=& -4(D_1^{(c)} + 2D_{12}^{(c)}) \\
A_{10} &=& 2(D_0^{(c)} - 2D_3^{(c)} - 4D_{12}^{(c)}) \\
A_{11} &=& -2 \left[ D_0^{(c)} + 4D_2^{(c)} - 2D_3^{(c)} + 4(D_{22}^{(c)}
 + D_{23}^{(c)}) \right] \\
A_{12} &=& -2 \left[ D_0^{(c)} - 2D_2^{(c)} - 4(D_{12}^{(c)} + D_{22}^{(c)}
 + D_{23}^{(c)}) \right] \\
A_{13} &=& -4(D_0^{(c)}+ D_2^{(c)} - D_3^{(c)} - 2D_{12}^{(c)}) \\
A_{14} &=& 4 \left[D_2^{(c)} - D_3^{(c)} + 2(D_{22}^{(c)}
 + D_{23}^{(c)}) \right] \\
\end{eqnarray*}

The overall factor for diagram $(d)$ is
$\frac{1}{8\pi^2}G^{f\bar{f}H}(G_L^{\bar{f}fZ} - G_R^{\bar{f}fZ})
G^{\bar{f}fg_1}G^{\bar{f}fg_2}m_f$,
and the coefficients are
\begin{eqnarray*}
A_1 &=& -2C_0^{(d)}+4C_2^{(d)}+(m_Z^2-\hat{u}-4)D_0^{(d)}+2\hat{s} D_3^{(d)} \\
 &&\hspace{3cm}+8D_{00}^{(d)}+4(\hat{u}-m_Z^2)D_{12}^{(d)}+4\hat{u}D_{22}^{(d)}
 +4(\hat{u} - \hat{t})D_{23}^{(d)} \\
A_2 &=& 4C_2^{(d)} + (m_Z^2 - \hat{t}) \left[ D_0^{(d)} - 2D_2^{(d)}
 - 4(D_{12}^{(d)} + D_{22}^{(d)}) \right]
 + 8D_{00}^{(d)} - 4\hat{s}(D_{13}^{(d)} + D_{23}^{(d)}) \\
A_3 &=& 2C_0^{(d)} + 4C_1^{(d)} - \hat{s} \left[ D_0^{(d)} - 2D_2^{(d)}
 - 4(D_{12}^{(d)} + D_{22}^{(d)} + D_{22}^{(d)}) \right] \\
A_4 &=& 4 \left[ D_3^{(d)} + 2(D_{13}^{(d)} + D_{23}^{(d)}) \right] \\
A_5 &=& -4 \left[ D_0^{(d)} + D_1^{(d)} + D_2^{(d)} - 2(D_{13}^{(d)}
 + D_{23}^{(d)}) \right] \\
A_6 &=& -4(D_3^{(d)} + 2D_{23}^{(d)}) \\
A_7 &=& A_9 = -4(D_2^{(d)} + 2D_{22}^{(d)}) \\
A_8 &=& 4 \left[ D_1^{(d)} + D_2^{(d)} + 2(D_{12}^{(d)}
 + D_{22}^{(d)}) \right] \\
A_{10} &=& -2 \left[ D_0^{(d)} - 4(D_2^{(d)} + D_{12}^{(d)}
 + D_{22}^{(d)}) \right] \\
A_{11} &=& 2(D_0^{(d)} + 4D_{23}^{(d)}) \\
A_{12} &=& 2 \left[ D_0^{(d)} - 2D_2^{(d)} - 4(D_{12}^{(d)} + D_{22}^{(d)}
 + D_{23}^{(d)}) \right] \\
A_{13} &=& 2 \left[ D_0^{(d)} - D_1^{(d)} - 2D_2^{(d)} - 4(D_{12}^{(d)}
 + D_{22}^{(d)}) \right] \\
A_{14} &=& 4(D_2^{(d)} - 2D_{23}^{(d)}) \\
\end{eqnarray*}

For diagram $(e)$, the overall factor is
$\frac{1}{8\pi^2}G^{f\bar{f}H}(G_L^{\bar{f}fZ} - G_R^{\bar{f}fZ})
G^{\bar{f}fg_1}G^{\bar{f}fg_2}m_f$,
and coefficients are
\begin{eqnarray*}
A_1 &=& -2C_0^{(e)} + (\hat{u} - m_Z^2)D_0^{(e)} + 2(m_Z^2 + \hat{u}) D_1^{(e)}
 + 4\hat{u} D_2^{(e)} + (m_Z^2 + \hat{s} - \hat{u})D_3^{(e)} + 8D_{00}^{(e)} \\
 && \hspace{1cm} + 4m_Z^2D_{11}^{(e)} + 4(m_Z^2 + \hat{u})(D_{12}^{(e)}
 + D_{22}^{(e)})
 + 4(\hat{u} - m_Z^2)D_{13}^{(e)} + \left[ 4(\hat{u} - m_Z^2) - \hat{s} \right]
 D_{23}^{(e)} \\
A_2 &=& -2C_0^{(e)} + 8D_{00}^{(e)}
 + 2(\hat{t} - m_Z^2) \left[ D_1^{(e)} + D_2^{(e)} + 2(D_{12}^{(e)}
 + D_{22}^{(e)}) \right] - 4\hat{s} D_{23}^{(e)} \\
A_3 &=& 2\hat{s}\left[D_1^{(e)}+3D_2^{(e)}+2(D_{13}^{(e)}+D_{23}^{(e)})\right]
 +2(2\hat{s}-m_Z^2)D_{12}^{(e)}+(\hat{t}-m_Z^2+5\hat{s})D_{22}^{(e)} \\
A_4 &=& 4(D_3^{(e)} + 2D_{23}^{(e)}) \\
A_5 &=& 4(D_2^{(e)} + 2D_{23}^{(e)}) \\
A_6 &=& -4 \left[ D_3^{(e)} + 2(D_{13}^{(e)} + D_{23}^{(e)}) \right] \\
A_7 &=& -4 \left[ D_0^{(e)} + 3(D_1^{(e)} + D_2^{(e)}) + 2(D_{11}^{(e)}
 + 2D_{12}^{(e)} + D_{22}^{(e)}) \right] \\
A_8 &=& 4 \left[ D_2^{(e)} + 2(D_{12}^{(e)} + D_{22}^{(e)}) \right] \\
A_9 &=& -4 \left[ D_1^{(e)} + D_2^{(e)} + 2(D_{11}^{(e)} + 2D_{12}^{(e)}
 + D_{22}^{(e)}) \right] \\
A_{10} &=& 8(D_2^{(e)} + D_{12}^{(e)} + D_{22}^{(e)}) \\
A_{11} &=& 8(D_{13}^{(e)} + D_{23}^{(e)}) \\
A_{12} &=& -2 \left[ D_0^{(e)} + 2D_1^{(e)} + 6D_2^{(e)}
 + 4(D_{12}^{(e)} + D_{22}^{(e)} + D_{13}^{(e)} + D_{23}^{(e)}) \right] \\
A_{13} &=& -4 \left[ D_1^{(e)} + 3D_2^{(e)} + 2(D_{12}^{(e)}
 + D_{22}^{(e)}) \right] \\
A_{14} &=& -2 \left[ D_0^{(e)} + 2(D_1^{(e)} + D_2^{(e)}) + 4(D_{13}^{(e)}
 + D_{23}^{(e)}) \right]
\end{eqnarray*}

\newpage

\begin{thebibliography}{99}

\bibitem{lhc}
F.~Gianotti, M.~L.~Mangano and T.~Virdee, CERN-TH/2002-078, \eprint{hep-ph/0204087}.

\bibitem{massh}
T.~Hambye and K.~Riesselmann, Phys. Rev. \textbf{D55}, 7255 (1997).

\bibitem{pdg}
Particle Data Group (K.~Hagiwara \textit{et al.}), Phys. Rev.
\textbf{D66}, 010001 (2002).

\bibitem{thdm}
N.~G.~Deshpande and E.~Ma, Phys. Rev. \textbf{D18}, 2574 (1978);
H.~Geogi, Hadronic Journal \textbf{1}, 155 (1978);
H.~E.~Haber, G.~L.~Kane and T.~Sterling, Nucl. Phys. \textbf{B161},
493 (1979);
J.~F.~Donoghue and L.~F.~Li, Phys Rev. \textbf{D19}, 945 (1979);
L.~F.~Abbott, P.~Sikivie and M.~B.~Wise, Phys. Rev. \textbf{D21},
1393 (1980);
B.~McWilliams and L.~F.~Li, Nucl. Phys. \textbf{B179}, 62 (1981).

\bibitem{mssm}
H.~E.~Haber and G.~L.~Kane, Phys. Rep. \textbf{117}, 75 (1985).

\bibitem{massh0}
H.~E.~Haber and R.~Hempfling, Phys. Rev. Lett. \textbf{66}, 1815 (1991);
Y.~Okada, M.~Yamaguchi and T.~Yanagida, Prog. Theor. Phys. \textbf{85}, 1
(1991);
J.~Ellis, G.~Ridolfi and F.~Zwirner, Phys. Lett. \textbf{B257}, 83 (1991).

\bibitem{detect}
A.~Djouadi, \eprint{CERN TH/2003-043}, \eprint{hep-ph/0303097};
M.~Dittmar, talk given at WHEPP 1999, Pramana 55, 151 (2000);
F.~Gianotti, talk given at the LHC Committee Meeting, CERN, 5/7/2000.

\bibitem{gg2h}
H.~Georgi \textit{et al.}, Phys. Rev. Lett. \textbf{40}, 692 (1978).

\bibitem{gg2hnlo}
M.~Spira, A.~Djouadi, D.~Graudenz and P.~Zerwas, Phys. Lett. \textbf{B318},
347 (1993); Nucl. Phys. \textbf{B453}, 17 (1995);
S.~Dawson, A.~Djouadi and M.~Spira, Phys. Rev. Lett. \textbf{77}, 16 (1996);
A.~Djouadi and M.~Spira, Phys. Rev. \textbf{D62}, 014004 (2000).

\bibitem{gg2hnnlo}
R.~V.~Harlander and W.~Kilgore, Phys. Rev. Lett. \textbf{88}, 201801 (2002);
J. High Energy Phys. \textbf{0210}, 017 (2002);
C.~Anastasiou and K.~Melnikov, Nucl. Phys. \textbf{B646}, 220 (2002);
Phys. Rev. \text{D67}, 037501 (2003);
V.~Ravindran, J.~Smith and W.~L.~van~Neerven, Nucl. Phys. \textbf{B665},
325 (2003).

\bibitem{gg2hresum}
S.~Catani, D.~de~Florian, M.~Grazzini and P.~Nason,
J. High Energy Phys. \textbf{0307}, 028 (2003);
A.~Kulesza, G.~Sterman and W.~Vogelsang, \eprint{hep-ph/0309264}.

\bibitem{vv2h}
R.~N.~Cahn and S.~Dawson, Phys. Lett. \textbf{B136}, 196 (1984);
G.~Altarelli, B.~Mele and F.~Pitolli, Nucl. Phys. \textbf{B287},
205 (1987);
T.~Han, G.~Valencia and S.~Willenbrock, Phys. Rev. Lett. \textbf{69},
3274 (1992).

\bibitem{v2vh}
S.~L.~Glashow, D.~V.~Nanopoulos and A.~Yildiz, Phys. Rev. \textbf{D18},
1724 (1978);
R.~Kleiss, Z.~Kunszt and W.~J.~Stirling, Phys. Lett. \textbf{B253},
269 (1991);
T.~Han and S.~Willenbrock, Phys. Lett. \textbf{B273}, 167 (1991).

\bibitem{tth}
Z.~Kunszt, Nucl. Phys. \textbf{B247}, 339 (1984);
W.~Beenakker \textit{et al.}, Phys. Rev. Lett. \textbf{87}, 201805 (2001);
Nucl. Phys. \textbf{B653}, 151 (2003);
S.~Dawson \textit{et al.}, Phys. Rev. Lett. \textbf{87}, 201804 (2001);
Phys. Rev. \textbf{D67}, 071503 (2003).

\bibitem{hpair}
A.~A.~Barrientos~Bendez\'{u} and B.~A.~Kniehl, Phys. Rev.
\textbf{D64}, 035006 (2001).

\bibitem{tevatron}
M.~Carena \textit{et al.}, hep-ph/0010338.

\bibitem{invisible}
D.~Choudhury and D.~P.~Roy, Phys. Lett. \textbf{B322}, 368 (1994);
R.~M.~Godbole, M.~Guchait, K.~Mazumdar, S.~Moretti,
and D.~P.~Roy, Phys. Lett. \textbf{B571}, 184 (2003).

\bibitem{kniehl}
B.~A.~Kniehl, Phys. Rev. \textbf{D42}, 2253 (1990).

\bibitem{feynarts}
T.~Hahn, Comp. Phys. Comm. \textbf{140}, 418 (2001).

\bibitem{alphas}
S.~G.~Gorishny \textit{et al.}, Mod. Phys. Lett. \textbf{A5}, 2703 (1990);
 Phys. Rev. \textbf{D43}, 1633 (1991);
A.~Djouadi, M.~Spira and P.~M.~Zerwas, Z. Phys. \textbf{C70}, 427 (1996);
A.~Djouadi, J.~Kalinowski and M.~Spira, Comput. Phys. Commun. \textbf{108},
 56 (1998).

\bibitem{spira}
M.~Spira, Fortschr. Phys. \textbf{46}, 203 (1998).

\bibitem{cteq}
J.~Pumplin, D.~R.~Stump, J.~Huston, H.~L.~Lai, P.~Nadolsky
 and W.~K.~Tung, J. High Energy Phys. \textbf{0207}, 012 (2002).

\bibitem{isajet}
H.~Baer \textit{et al}.,
Manual of ISAJET version 7.69 at \url{http://www.phy.bnl.gov/~isajet};
\eprint{hep-ph/0001086} (1999).

\bibitem{loop}
G.~Passarino and M.~Veltman, Nucl. Phys. \textbf{B160}, 151 (1979);
A.~Denner, Fortschr. Phys. \textbf{41}, 4 (1993).

\end{thebibliography}
\end{document}